\documentclass{elsart}

\usepackage{amssymb,amsmath,mathrsfs,eufrak}
\newtheorem{exm}{Example}[section]

\newcommand{\mP}{\mathcal{P}}

\newcommand{\N}{\mathcal{N}}

\newcommand{\ar}{\rightarrow}
\newcommand{\la}[1]{\it {#1}\rm}

\newcommand{\mT}{\mathcal{T}}
\newcommand{\mL}{\mathcal{L}}

\newcommand{\mV}{\mathcal{V}}

\newcommand{\df}[1]{\begin{defn} #1 \end{defn}}

\newcommand{\lm}[1]{\begin{lem} #1 \end{lem}}
\newcommand{\te}[1]{\begin{thm} #1 \end{thm}}
\newcommand{\ex}[1]{\begin{exm} #1 \end{exm}}

\newcommand{\F}{{\mathsf F }}
\newcommand{\TT}{{\mathsf T }}

\begin{document}

\begin{frontmatter}



\title{Clone Theory and Algebraic Logic}
%\large{(Research Announcement)}


{\large Zhaohua Luo}
\author{}

%\small{12/1/2007}

% \address{Geometry.net, 1800 W. Hillcrest 235, Newbury Park, CA, 91320}
 \ead{zluo@azd.com}
\ead[url]{http://www.algebraic.net/cag}

\begin{abstract}
%Text of abstract

The concept of a clone is central to many branches of mathematics, such as universal algebra, algebraic logic, and lambda calculus. Abstractly a clone is a category with two objects such that one is a countably infinite power of the other. Left and right algebras over a clone are covariant and contravariant functors from the category to that of sets respectively. In this paper we show that first-order logic can be studied effectively using the notions of right and left algebras over a clone. It is easy to translate the classical treatment of logic into our setting and prove all the fundamental theorems of first-order theory algebraically.

\end{abstract}

%\begin{keyword}
%Clone \sep Universal Algebra \sep First-Order Theory \sep Lambda
%Calculus \sep Polyadic Algebra
%\PACS
%\end{keyword}
\end{frontmatter}

% main text
\section*{Introduction}

The theory of clones has been introduced in two previous papers \cite{luo:1} and \cite{luo:2}. In the present paper we are mainly concerned with the applications of clone theory to mathematical logic, as an extension of the last two sections of \cite{luo:2}.

A clone is a right act over a monoid which is the countably infinite power of the right act. The concept of a clone is central to many branches of mathematics, such as universal algebra, algebraic logic, and lambda calculus. Abstractly a clone is a category with two objects such that one is the countably infinite power of the other. A right algebra over a clone corresponds to a contravariant set functor, and a left algebra over a clone corresponds to a covariant set functor preserving the countably infinite power.

In this paper we show that first-order logic can be studied effectively using the notions of right and left algebras over a clone. Let $\mL$ be a first-order language. The terms of $\mL$ form a free clone $F(X)$ generated by the function symbols, and the formulas of $\mL$ form a free right algebra $P(\mL)$ over $F(X)$ generated by the predicate symbols. A model for $\mL$ is then determined by a left algebra over $F(X)$. It is easy to translate the classical treatment of logic into our setting and prove all the fundamental theorems of first-order theory algebraically.

\section{Proposition Algebras}
A \la{proposition algebra} is an algebra $(P, \wedge, \neg)$, where $P$ is a set, and

$\wedge: P \times P \ar P$,

$\neg: P \ar P$,

are operations on $P$.

If $P$ is a proposition algebra for any $p, q \in P$ let

$p \vee q = \neg ((\neg p) \wedge (\neg q))$.

$p \ar q = (\neg p) \vee q$.

$p \leftrightarrow q = (p \ar q) \wedge (q \ar p)$.

$\F_P = \{p \wedge (\neg p) | p \in P\}$.

$\TT_P = \{p \vee (\neg p) | p \in P\}$.

A \la{(truth) valuation} (or \la{proposition valuation}) of a proposition algebra $P$ is a subset $V$ of $P$ such that for any $p, q \in P$ we have

$p \in V$ iff $\neg p \ne V$.

$p \wedge q \in V$ iff $p \in V$ and $q \in V$.

Denote by $Val(P)$ the intersection of all valuations of $P$; an element of $Val(P)$ is called a \la{logically valid element} of $P$. Note that $\TT_P \subset Val(P)$. If $P, Q$ are proposition algebras and $\phi: P \ar  Q$ is a homomorphism of proposition algebras then $\phi(Val(P)) \subset Val(Q)$ and $\phi(\TT_P) \subset \TT_Q$.

A subset $F$ of $P$ is called \la{MP-closed} (MP for \la{Modus Ponens}) if $p, p \ar q \in F$ implies that $q \in F$ for any $p, q \in P$.

Let $p, q, r \in P$. Then each of the following elements is logically valid, called an \la{axiom} of $P$:

(A1) $p \ar (p \wedge p)$.

(A2) $p \wedge q \ar p$.

(A3) $(p \ar q) \ar (\neg (q \wedge r) \ar \neg (r \wedge p))$.

A \la{(proposition) filter} of a proposition algebra $P$ is a MP-closed subset $F$ of $P$ containing every axiom of $P$.

The class of filters is closed under intersection.

Suppose $T$ is any subset of $P$.

Denote by $Con(T)$ the intersection of all valuations of $P$ containing $T$; an element of $Con(T)$ is called a \la{consequence} of $T$.

Denote by $Ded(T)$ the intersection of filters of $P$ containing $T$.

If $S, T$ are two subsets of $P$ we write

(i) $T \models S$ if $Con(T) \supseteq S$, and

(ii) $T \vdash S$ if $Ded(T) \supseteq S$.

If $T = \emptyset$ we write $\models S$ (resp. $\vdash S$) instead of $\emptyset \models S$ (resp. $\emptyset \vdash S$).

Let $p \in P$ and $T \subset P$. A \la{proof of $p$ from $T$} is a finite sequence $p_1, p_2, ..., p_n$ of elements of $P$ such that $p = p_n$ and for each $i \le n$, either

(i) $p_i$ is an axiom, or

(ii) $p_i \in T$, or

(iii) for some $j, k < i$ we have $p_k = (p_j \ar p_i)$.

We say that $p$ is a \la{deduction} from $T$, or $p$ is \la{provable} from $T$, if there exists a proof of $p$ from $T$.

\lm{$T \vdash p$ iff $p$ is a deduction from $T$.}

A subset $T$ of $P$ is called \la{inconsistent} if there is $p \in P$ such that both $p, \neg p \in Ded(T)$; otherwise we say that $p$ is \la{consistent}.

A filter is  \la{maximal} if it is a proper subset that is not a proper subset of any other proper filter of $P$.

\lm{1. A filter $F$ is maximal iff $p \in F \Leftrightarrow \neg p \notin F$ for any $p \in P$.

2. Any consistent subset of $P$ is contained in a maximal filter of $P$.}

\te{(Completeness Theorem for Proposition Algebras) Let $F$ be a subset of a proposition algebra $P$.

 (i) $F$ is a filter of $P$ iff it is the intersection of all valuations of $P$ containing $F$.

(ii) $F$ is a maximal filter iff it is a valuation of $P$.

(iii) $T \models S$ iff $T \vdash S$ for any two subsets $T, S$ of $P$.}

A \la{Boolean algebra} is a proposition algebra $(P, \wedge, \neg)$ such that for any $p, q \in P$, if $p \leftrightarrow q$ is valid (i.e. $p \leftrightarrow q$ is contained in any truth valuation of $P$) then $p = q$.

Algebraically a \la{Boolean algebra} can be defined as a proposition algebra $(P, \wedge, \neg)$ satisfying the following conditions for any  $p, q, r \in P$:

(i) $p \wedge (q \wedge r) = (p \wedge q) \wedge r$.

(ii) $p \wedge q = q \wedge p$.

(iii) If $p \wedge (\neg q) = r \wedge (\neg r)$ then $p \wedge q  = p$.

(iv) If $p \wedge q = p$ then $p \wedge (\neg q) = r \wedge (\neg r)$.

Suppose $P$ is a Boolean algebra. Then $\F_P$ and $\TT_P$ are singletons. Let $\F_P = \{0\}$ and $\TT_P = \{1\}$, Then $(P, \vee, \wedge, 0, 1)$ is a complemented   distributive lattice with the partial order on $P$ defined by
\[p \le q \Leftrightarrow p \wedge q = p.\]
for all elements $p$ and $q$ in $P$.

\lm{A subset $F$ of a Boolean algebra $P$ is a filter iff the following conditions are satisfied:

(i) If $p, q \in F$ then $p \wedge q \in F$.

(ii) If $p \in F$ then $p \vee q \in F$ for any $q \in P$.

(iii) $1 \in F$.
}

A congruence $\sim$ on $P$ is called \la{regular} if the following conditions are satisfied:

1. $p \sim q$ for any $p, q \in \TT_P$.

2. The equivalence class $[\TT_P]$ determined by $\TT_P$ is a filter.

3. $p \sim q$ iff $p \leftrightarrow q \in [\TT_P]$ for any $p, q \in P$.

Any filter $F$ determines a unique regular congruence, denoted by $\sim_F$.

If $\sim$ is a regular congruence on a proposition algebra $P$ then the quotient algebra $P/\sim$ is a Boolean algebra. A congruence determined by a homomorphism $P \ar Q$ of proposition algebras is regular iff $Q$ is a Boolean algebra.

\df{If $P$ is a proposition algebra then the Boolean algebra $P/\sim_{Val(P)}$ is called the \la{Lindenbaum Boolean algebra} of $P$.
}

\ex{Let $2 = \{0, 1\}$ be the set of two elements $0, 1$. Define $\neg 1 =  0 \wedge 0 = 0 \wedge 1 = 1 \wedge 0 = 0$ and $\neg 0 = 1 \wedge 1 = 1$. Then $(2, \wedge, \neg)$ is a Boolean algebra. Any valuation $V$ of a proposition algebra $P$ determines a homomorphism $P \ar 2$ sending $V$ to $1$. Conversely, any such homomorphism arises in this way. }

\ex{
Let $X$ be a nonempty set of variables. Let $\mT(X)$ be the smallest set containing $X$ such that if $p, q \in \mT(X)$ then $\neg p, p \wedge q \in \mT(X)$. Then $\mT(X)$ is a free proposition algebra over $X$. A logically valid element of $\mT(X)$ is called a \la{tautology}. The Lindenbaum Boolean algebra of $\mT(X)$ is a free Boolean algebra over $X$.
}
\section{Clones}

A \la{monoid} is a set $G$ together with an element (identity) $1$ of $G$ and a multiplication $G \times G \ar G$ such that for any $u, v, w \in G$  we have

$u (v w) = (u  v) w$.

$1  u = u  1 = u$.

A \la{right act over a monoid $G$} is a set $P$ together with a multiplication $P \times G \ar P$ such that for any $p \in P$ and $u, v \in G$ we have

$(pu)v = p(uv)$.

$p1 = p$.

Let $\N$ be the set of positive integers. If $A$ is any nonempty set denote by $A^{\N}$ the set of   infinite sequences $[a_1, a_2, ...]$ of elements of $A$.

A \la{clone} is a nonempty set $A$ such that

(i) $A^{\N}$ is a monoid with an identity $[x_1, x_2, ...]$.

(ii) $A$ is a right act over $A^{\N}$.

(iii) $x_i[a_1, a_2, ...] = a_i$ for any $i > 0$.

Alternatively a clone can be defined as a set $A$ containing a set $X = \{x_1, x_2, ...\}$ of variables together with a multiplication $A \times A^{\N} \ar A$ such that for any $a, a_1, a_2, ..., b_1, b_2, ... \in A$ we have

(i) $(a[a_1, a_2, ...])[b_1, b_2, ...] = a[a_1[b_1, b_2, ...], a_2[b_1, b_2, ...], ...]$.

(ii) $a[x_1, x_2, ...] = a$.

(iii) $x_i[a_1, a_2, ...] = a_i$ for any $i > 0$.

\ex{$X = \{x_1, x_2, ...\}$ is a clone if we define $x_i[x_{k_1}, x_{k_2}, ...] = x_{k_i}$. It is the initial clone in the category of clones.}

Suppose $A$ is a clone.

A \la{right algebra over $A$} (or \la{right $A$-algebra}) is a right act $P$ over the monoid $A^{\N}$.

Suppose $P$ is a right algebra over $A$.

For any $p \in P$ let

$p^+ = p[x_2, x_3, ...]$,

 $p^- = p[x_1, x_1, x_2, x_3, ...]$.

 $p^* = p[x_2, x_2, x_3, x_4, ...]$.

 Then $(p^+)^- = p$ and $(p^-)^+ = p^*$

If $p \in P$ and $a_1, ... , a_n \in A$ we write $p[a_1, ..., a_n]$ as an abbreviation for $p[a_1, ..., a_{n-1}, a_n, a_n, a_n, a_n, ...]$.

We say a right $A$-algebra $P$ is \la{locally finite} if for any $p$ there is $n > 0$ (called a \la{finite rank of} $a$) such that $p = p[x_1, ..., x_n]$. An element $p \in P$ is called \la{closed} (or \la{with a finite rank} $0$) if $p[a_1, a_2, ...] = p$ for any $a_1, a_2, ... \in A$.

A \la{left algebra over a clone $A$}  (or \la{left $A$-algebra}) is a set $M$ together with a multiplication $A \times M^{\N} \ar M$ such that  for any $a, a_1, a_2, ... \in A$ and $m_1, m_2, ... \in M$ we have

(i) $(a[a_1, a_2, ...])[m_1, m_2, ...] = a[a_1[m_1, m_2, ...], a_2[m_1, m_2, ...], ...]$.

(ii) $x_i[m_1, m_2, ...] = m_i$ for any $i > 0$.

The class of left $A$-algebras is a variety, which is a finitary variety iff $A$ is locally finite. If $Y$ is any set then free left $A$-algebras $A(Y)$ over $Y$ exists.

Homomorphisms of clones, right algebras,  and left algebras over a clone are defined in an obvious way (cf. \cite{luo:2}).

Let $C$ be a concrete category over the category of sets (such as the of categories of sets, proposition algebras, Boolean algebras, or any variety).  A \la{transformation algebra over} a clone $A$ is an object $P$ of a concrete category $C$ together with a multiplication $P \times A^{\N} \ar P$ such that the following conditions are satisfied for any $p \in P$ and $a_1, a_2, ..., b_1, b_2, ... \in A$:

(T1) $(p[a_1, a_2, ...])[b_1, b_2, ...] = p[a_1[b_1, b_2, ...], a_2[b_1, b_2, ...], ...]$.

(T2) $p[x_1, x_2, ...] = p$.

(T3) The function $\phi_{[a_1, a_2, ...]}: P \ar P$ sending $p$ to $p[a_1, a_2, ...]$ is an endomorphism on $P$.

If $C$ is a finitary variety (of algebras) then (T3) has the following explicit form:

(T4) For any $n$-ary fundamental operation $f: P^n \ar P$ on $P$ we have $(f(p_1, ..., p_n))[a_1, a_2, ...] = f(p_1[a_1, a_2, ...], ..., p_n[a_1, a_2, ...])$.

Note that (T1) and (T2) imply that $P$ is a right $A$-algebra, called the \la{underlying right $A$-algebra of $P$}. We say a transformation algebra $P$ is \la{locally finite} if the underlying right $A$-algebra of  $P$ is locally finite.

A transformation algebra over $A$ in the category of sets, proposition algebras, Boolean algebras, ... is called a \la{transformation set, transformation proposition algebras, transformation Boolean algebra}, ... over $A$. Note that a transformation set over $A$ is just a right algebra over $A$. Thus a clone is a transformation set over itself.

An \la{abstract binding operation} on a transformation algebra $P$ over a clone $A$ is a function $\forall: P \ar P$ such that for any $p \in P$ and $a_1, a_2, ...  \in A$ we have
\[(\forall p)[a_1, a_2, ...] = \forall (p[x_1, a_1^+, a_2^+, ...]).\]
 If $\forall$ is an abstract binding operation, for any positive integer $i > 0$ the conventional \la{ $i$-th binding operation } $\forall x_i$ on $P$ is defined by \[\forall x_i.p = \forall(p[x_2, x_3, ..., x_{i-1}, x_i, x_1, x_{i+2}, ...]).\]
 If $p \in P$ and $n \ge 0$ let \[\forall^n p = \forall (\forall (....(\forall p)...)).\]We assume $\forall^0 p = p$. 
 
 \lm{Suppose $p$ is an element of $P$.
 
 1. If $p$ is closed then $\forall p$ is closed.
 
 2. If $p$ has a finite rank $n > 0$ then $\forall p$ has a finite rank $n-1$.
 
 3. If $p$ has a finite rank $n > 0$ thus $\forall^n p$ is closed.
 
 4. $\forall (p[x_1])$ is closed.
 
 5. $(\forall x_i.p)[x_1, x_2, ..., x_{i-1}, a, x_{i+1}, x_{i+2}, ...] = \forall x_i.p$ for any $a \in A$.
 
 6. If $p$ has a finite rank $n > 0$ then $\forall x_1.(...(\forall x_{n-1}.(\forall x_n. p))...)$ is closed.}
  
 This implies that if $P$ carries an abstract binding operation, then the set of closed elements of $P$ is not empty.

\ex{There is no abstract binding operation on the initial clone $X = \{x_1, x_2, ...\}$ because it has no closed element.}

If $\forall$ is an abstract binding operation on a transformation proposition algebra $P$ we define $\exists: P \ar P$ and $\exists x_i$ for every $i > 0$ by
\[\exists p = \neg (\forall (\neg p)).\]
\[\exists x_i.p = \exists(p[x_2, x_3, ..., x_{i-1}, x_i, x_1, x_{i+2}, ...]).\]Then $\exists$ is also an abstract binding operation on $P$.

We have \[\forall x_1.p = \forall (p[x_1, x_3, x_4, ...]) = (\forall p)^+.\]
 \[\exists x_1.p = \exists (p[x_1, x_3, x_4, ...]) = (\exists p)^+.\]
So \[\forall p = (\forall x_1.p)^-.\]
\[\exists p = (\exists x_1.p)^-.\]

\section{Predicate Algebras}

Let $A$ be a clone.

A \la{predicate (proposition) algebra} over $A$ is a transformation proposition algebra $P$ over $A$ together with an abstract binding operation $\forall$ on $P$; if an element $e \in P$ of rank $2$ is specified then we say that $P$ is a predicate algebra \la{with equality $e$}.

A \la{quantifier (Boolean) algebra} over $A$ is a transformation Boolean algebra $P$ over $A$ together with an abstract binding operation $\forall$ on $P$ satisfying the following conditions for any $p, q \in P$:

(Q1) $\forall(p \wedge q) = \forall p \wedge \forall q$.

(Q2) $(\forall  p)^+ = (\forall p)^+ \wedge p$.

(Q3) $\forall ( p^+) = p$.

An element $e \in P$ of rank $2$ is called an \la{equality} for a quantifier algebra $P$ if the following two conditions are satisfied:

(Q4) $e^* = 1$.

(Q5) $e \wedge p = e \wedge p^*$.

The axioms (Q1)-(Q5) are justified by the following observations:

1. Any abstract binding operation $\forall$ on a predicate algebra satisfying the axioms (Q1), Q(2) and (Q3) is unique if exists.

2. Any element $e$ in a quantifier algebra satisfying the axioms (Q4) and (Q5) is unique if exists.

3. There are plenty of concrete quantifier algebras (see Section \ref{sec:fun}).

We say a quantifier algebra $(P, \forall)$ is \la{nontrivial} if $0 \ne 1$.

We say a quantifier algebra $(P, \forall)$ is \la{simple} if $0 \ne 1$ and these are the only closed elements of $P$.

The class of predicate algebras (resp. quantifier algebras) over a clone forms a finitary variety.

In the same way we obtain the varieties of predicate (resp. quantifier) Post algebras, Heyting algebras, frames, etc.

One can show that the variety of locally finite quantifier (Boolean) algebras over the initial clone $X = \{x_1, x_2, ...\}$ is equivalent to the variety of locally finite polyadic algebras of countably infinite degree (cf. \cite{h:1}). Thus a quantifier algebra over an arbitrary clone may be viewed as a polyadic algebra with terms.

\section{\label{sec:fun}Models}

Let $A$ be a clone and let $M$ be a left algebra over $A$.

Suppose $B$ is a Boolean algebra. Let $B^{M^{\N}}$ be the set of functions from $M^{\N}$ to $B$.

For any $p, q \in B^{M^{\N}}$ and $a_1, a_2, ... \in A$ we define $\neg p, p \wedge g, p[a_1, a_2, ...], \forall p, e \in B^{M^{\N}}$ such that for any $m_1, m_2, ... \in M$ we have

$(\neg p)[m_1, m_2, ...] = \neg(p[m_1, m_2, ...])$.

$(p \wedge q)[m_1, m_2, ...] = p[m_1, m_2, ...] \wedge q[m_1, m_2, ...]$.

$(p[a_1, a_2, ...])[m_1, m_2, ...] = p(a_1[m_1, m_2, ...], a_2[m_1, m_2, ...], ...]$.

$(\forall p)[m_1, m_2, ...] = \bigwedge \{p[m, m_1, m_2, ...] :  m \in M\}$ if the right side meet exists.

$e[m_1, m_2, ...] = 1$ if $m_1 = m_2$ and $e[m_1, m_2, ...] = 0$ otherwise.

Then $(B^{M^{\N}}, \wedge, \neg, e)$ is a transformation Boolean algebra with equality $e$ over $A$.

If $B$ is a complete Boolean algebra then $\forall$ is an abstract binding operation defined everywhere on $B^{M^{\N}}$, and $\mP_B(M) = (B^{M^{\N}}, \forall, e)$ is a  quantifier algebra with equality. If $2 = \{0, 1\}$ then $\mP_2(M) = 2^{M^{\N}}$ is called the \la{classical functional quantifier algebra determined by $M$}.

If $B$ is any Boolean algebra  by a \la{functional quantifier algebra over $B$} we mean a pair $(P, M)$ where $M$ is a left algebra over $A$, $P$ is a subalgebra of the  transformation Boolean algebra $B^{M^{\N}}$ over $A$ such that if $p \in P$ then $\forall p$ is defined and $\forall p \in P$.

Let $P$ be a predicate algebra over a clone $A$.

A \la{model over $B$ for $P$ } is a pair $M = (M,  \Omega)$ where $M$ is a nonempty left algebra over $A$, and $\Omega: P \times M^{\N} \ar B$ is a multiplication such that for any $p, p_1, p_2 \in P$, $m_1, m_2, ... \in M$ and $a_1, a_2, ... \in A$ we have

M1. $(\neg p)[m_1, m_2, ...] = \neg (p[m_1, m_2, ..])$.

M2. $(p_1 \wedge p_2)[m_1, m_2, ...] = p_1[m_1, m_2, ...] \wedge p_2[m_1, m_2, ...]$.

M3. $(p[a_1, a_2, ...])[m_1, m_2, ...] = p[a_1[m_1, m_2, ...], a_2[m_1, m_2, ...], ...])$.

M4. $(\forall p)[m_1, m_2, ...] = \bigwedge \{p[m, m_1,, m_2, ...] | m \in M\}$ (i.e. the right side meet exists which equals the left side).

If $P$ is a predicate algebra with equality $e$ we say a model $(M, \Omega)$ for $P$ is a \la{model preserving equality} if we have

M5. $e [m_1, m_2, m_3, ... ] = 1$ iff $m_1 = m_2$.

If $(M, \Omega)$ is a model over $B$ for $P$ then $\Omega$ induces two mappings \[\Omega^*: P \ar B^{M^{\N}}.\] \[\Omega_*: M^{\N} \ar B^P.\] where  $B^P$ is the set of homomorphisms of Boolean algebras from $P$ to $B$.

First we consider the mapping $\Omega^*$. The image of $P$ under $\Omega^*$ is a functional quantifier algebra over $B$ and $\Omega^*$ induces a homomorphism of predicate algebras from $P$ to $\Omega^*(P)$; we say $(M, \Omega)$ is a \la{faithful model} if $\Omega^*$ is injective. Conversely, any homomorphism of predicate algebras from $P$ to a  functional quantifier algebra over $B$ defines a model over $B$ for $P$.

\te{\label{te:cay} (Cayley's Theorem for Quantifier Algebras) If $P$ is a locally finite quantifier algebra over a locally finite clone $A$ then $A$ together with the canonical multiplication $P \times A^{\N} \ar P$ is a faithful model for $P$ over the Boolean algebra $P$, which is called the canonical model for $P$.}

Next we study the mapping $\Omega_*$. Since $P$ is a right act over $A^{\N}$, $B^P$ is a left act over $A^{\N}$. Also  $M^{\N}$ is a left act over $A^{\N}$. Clearly $\Omega_*$ is a homomorphism of left acts over $A^{\N}$ by conditions (M1) - (M3). Thus a model of $P$ over $B$ is determined by a left algebra $M$ over $A$ together with a homomorphism of left $A^{\N}$-acts from $M^{\N}$ to $B^P$ satisfying the condition (M4). 

Note that if $B = 2$ then $2^P$ is the Stone space for the Boolean algebra $P$. A model over $2$ for $P$ is called a \la{classical model} for $P$. Note that a multiplication $\Omega: P \times M^{\N} \ar 2$ is uniquely determined by the subset $U = \Omega^{-1}(1)$ of $P \times M^{\N}$. Thus a classical model for $P$  may be defined as a left algebra $M$ over $A$ together with a subset $U$ of $P \times M^{\N}$.

Suppose $(M, \Omega)$ is a classical model for $P$.

Let $\mV(M) = \{p \in P \ | \ p[m_1, m_2, ...] = 1 \ \text{ for any } \ m_1, m_2, ... \in M \}$.

Let $\mV_0(M) = P_0 \cap \mV(M)$, where $P_0$ is the set of closed elements of  $P$.

If $m_1, m_2, ... \in M$ let $\mV_{m_1, m_2, ...}(M) = \{p \in P \ | \ p[m_1, m_2, ...] = 1\}$.

A subset $V$ of a predicate algebra $P$ over $A$ is called a \la{global valuation} if there is a classical model $(M, \Omega)$ for $P$ such that $V = \mV(M)$.

A subset $U$ of closed elements of a predicate algebra $P$ over $A$ is called a \la{closed valuation} if there is a global valuation $V$ of $P$ such that $U = V \cap P_0$.

A subset $V$ of a predicate algebra $P$ over $A$ is called a \la{local valuation} if there is a classical model $(M, \Omega)$ for $P$ and an sequence $m_1, m_2, ... \in M$ such that $V = \mV_{m_1, m_2, ...}(M) $.

Note that any global valuation is an intersection of local valuations.

Let $Val(P)$ be the intersection of all global (or local) valuations of $P$; an element of $Val(P)$ is called a \la{logically valid element}  of $P$.

If $T$ is any subset of a predicate algebra $P$ over a clone $A$ we denote by $Con_l(T)$ (resp. $Con_g(T)$) the intersection of all the local valuations (resp. global valuations) containing $T$.

If $T, S$ are subsets of $P$ we write

$T \models_l S$ if $Con_l(T) \supseteq S$,

$T \models_g S$ if $Con_g(T) \supseteq S$,

If $T = \emptyset$ we write $\models_l S$ (or $\models_g S$)  instead of $\emptyset \models_l S$ (or $\emptyset \models_g S$).

\section{Filters}

Let $A$ be a clone. Let $P$ be a predicate algebra over $A$.

Suppose $F$ is a subset of $P$. We say $F$ is \la{closed under substitution} if $p \in F$ implies that $p[a_1, a_2, ...] \in F$ for any $a_1, a_2, ... \in A$. We say $F$ is \la{closed under generalization} if $p \in F$ implies that $\forall p \in F$. Recall that $F$ is \la{MP-closed} if $p, p \ar q \in F$ implies that $q \in F$ for any $p, q \in P$.

A subset $F$ of $P$ is called \la{globally closed} if the following conditions are satisfied:

(i) $F$ is MP-closed.

(ii) $F$ is closed under generalization.

(iii) $F$ is closed under substitution.

The class of globally closed subsets of $P$ is closed under intersection.

\lm{1. Any global valuation of $P$ is globally closed.

2. Any intersection of global valuations of $P$ is globally closed.

3. The set of logically valid elements of $P$ is globally closed.

}

Let $p, q, r \in P$ and $a_1, a_2, ... \in A$. Then each of the following elements is logically valid, called a \la{prime axiom}:

(A1) $p \ar (p \wedge p)$.

(A2) $p \wedge q \ar p$.

(A3) $(p \ar q) \ar (\neg (q \wedge r) \ar \neg (r \wedge p))$.

(A4) $\forall (p \ar q) \ar ((\forall p) \ar (\forall q))$,

(A5) $(\forall p)[a_2, a_3, ...]  \ar p[a_1, a_2, ...]$,

(A6) $p \ar \forall (p^+)$.

If $P$ has an equality $e$ then each of the following elements is also called a \la{prime axiom}:

(A7) $e[x_i, x_i]$ for any $i > 0$.

(A8) $e[a_1, a_2] \wedge p[a_1, a_2, ...] \ar p[a_2, a_2, a_3, ...]$ for any $a_1, a_2, ... \in A$.

Note that the set of prime axioms of $P$ is closed under substitution.

If $p$ is a prime axiom then  $\forall^n p$ is logically valid for any integer $n \ge 0$, called an \la{axiom} (we assume $\forall^0 p = p$).

Note the set of axioms of $P$ is closed under substitution and generalization.

\df{A subset $F$ of a predicate algebra $P$  is called a global filter of $P$ if the following conditions are satisfied:

(i) $F$ contains every prime axiom of $P$.

(ii) $F$ is globally closed.
}

\df{A subset $F$ of a predicate algebra $P$  is called a local filter of $P$ if the following conditions are satisfied:

(i) $F$ contains every axiom of $P$.

(ii) $F$ is MP-closed.
}

Alternatively, one can define a local filter of $P$ as a MP-closed subset of $P$ containing a global filter.

Every global filter of $P$ is a local filter, and every local filter is a proposition filter of the proposition algebra $P$.

The class of global filters (resp. local filters) of $P$ is closed under intersection.

Suppose $T$ is any subset of $P$. Denote by $Ded_l(T)$ (resp. $Ded_g(T)$) the intersection of all the local filters (resp. global filters) containing $T$, which is called the \la{local filter} (resp. \la{global filter}) \la{ generated by} $T$.

If $T, S$ are subsets of $P$ we write

$T \vdash_l S$ if $Ded_l(T) \supseteq S$,

$T \vdash_g S$ if $Ded_g(T) \supseteq S$.

If $T = \emptyset$ we write $\vdash_l S$ (or $\vdash_g S$)  instead of $\emptyset \vdash_l S$ (or $\emptyset \vdash_g S$).

A subset $T$ of $P$ is called \la{globally inconsistent} if there is $p \in P$ such that both $p, \neg p \in Ded_g(T)$; otherwise we say that $p$ is \la{globally consistent}.

A subset $T$ of $P$ is called \la{locally inconsistent} if there is $p \in P$ such that both $p, \neg p \in Ded_l(T)$; otherwise we say that $p$ is \la{locally consistent}.

A predicate algebra $P$ is \la{consistent} if the empty set is globally consistent (or equivalently, $P$ has a proper global filter).

Suppose $p$ is an element of $P$ and $T$ is a subset of $P$.

A \la{global proof of $p$ from $T$} is a finite sequence $p_1, p_2, ..., p_n$ of elements of $P$ such that $p = p_n$ and for each $i \le n$, either

$p_i$ is an axiom of $P$, or

$p_i \in T$, or

$p_i = p_j[a_1, a_2, ...]$ for some $j < i$ and $a_1, a_2, ... \in A$, or

$p_i = \forall p_j$ for some $j < i$, or

$p_k = (p_j \ar p_i)$ for some $j, k < i$.

We say that $p$ is a \la{global deduction} from $T$, or $p$ is \la{globally provable} from $T$, if there exists a proof of $p$ from $T$.

A \la{local proof of $p$ from $T$} is a finite sequence $p_1, p_2, ..., p_n$ of elements of $P$ such that $p = p_n$ and for each $i \le n$, either

$p_i$ is an axiom of $P(\mL)$, or

$p_i \in T$, or

$p_k = (p_j \ar p_i)$ for some $j, k < i$.

We say that $p$ is a \la{local deduction} from $T$, or $p$ is \la{locally  provable} from $T$, if there exists a local proof of $p$ from $T$.

\lm{1. $T \vdash_g p$ iff $p$ is a global deduction from $T$.

2. $T \vdash_l p$ iff $p$ is a local deduction from $T$.
}

A local (resp. global filter) $F$ is \la{maximal} if it is a proper subset that is not a proper subset of any other proper local (resp. global) filter of $P$.

Any global filter $F$ of $P$ determines a congruence $\sim_F$ on $P$ by
\[p \sim_F q \Leftrightarrow p \leftrightarrow q \in F. \]
A subset $F$ of $P$ is a global filter iff there is a homomorphism  $\phi: P \ar Q$ from $P$ to a quantifier algebra over $A$ such that $F = \phi^{-1}(1)$.

\df{A theory of a predicate algebra $P$ is a set $T$ of elements  of $P$; we say a theory $T$ is complete if $Ded_g(T)$ is maximal.}

If $T$ is a theory of $P$ then the quantifier algebra $Lin(T) = P/\sim_{Ded_g(T)}$ is called the \la{Lindenbaum algebra} of the theory $T$. If $T = \emptyset$ we call $Lin(\emptyset)$ the \la{Lindenbaum algebra}  of $P$, which is denoted by $Lin(P)$.

\lm{A local (resp. global) filter $F$ is \la{maximal} iff for any element (resp. closed element) $p \in P$ we have \[p \in F\Leftrightarrow \neg p \notin F.\]
}
\lm{Suppose $A$ is a locally finite clone and $P$ is a locally finite predicate algebra over $A$.

1. A local filter of $P$ is global iff it is generated by a set of closed elements.

2. The  lattice of proposition filters of $P_0$ is isomorphic to the lattice of global filters of $P$.

3. A global filter $F$ of $P$ is maximal iff $F \cap P_0$ is a maximal proposition filter of $P_0$.

4. A global filter $F$ of $P$ is maximal iff the quotient algebra $P/\sim_F$ is a simple quantifier algebra.

5. Any global filter $F$ of $P$ is the intersection of all maximal global filters of $P$ which contains $F$.
}

\lm{1. A quantifier algebra is consistent if it is nontrivial.

2. A closed subset of a quantifier algebra $T$ is globally consistent iff it has the finite meet property, i.e. whenever $p_1, ..., P_n \in T$ we have $p_1 \wedge... p_n \ne 0$.
}

\section{First-Order Algebras}

A \la{type} is a set of symbols such that each symbol has a non-negative integer (called \la{arity}) assigned to it.

Let $X = \{x_1, x_2, ...\}$ be a set of variables. If $F$ is a type we let $F(X)$ be the smallest set such that

1. $X \subset F(X)$.

2. If $f \in F$ is an $n$-ary symbol and $t_1, ..., t_n \in F(X)$ then $f[t_1, ..., t_n] \in F(X)$.

Definite $F(X) \times F(X)^{\N} \ar F(X)$ inductively:

$x_i[s_1, s_2, ...] = s_i$.

$(f[t_1, ..., t_n])[s_1, s_2, ... ] = f[t_1[s_1, s_2, ...], ..., t_n[s_1, s_2, ...]]$.

Then $F(X)$ is a locally finite clone; each expression in $F(X)$ is called a \la{term} over $F$ in $X$.

The locally finite clone $F(X)$ has the following universal property.

\lm{Suppose $B$ is a clone. Suppose $\phi: F \ar B$ is a function such that for each $n$-ary $f \in F$ the element $\phi(f)$ has a finite rank $n$. Then $\phi$ extends uniquely to a homomorphism of clones from $F(X)$ to $B$.}

Let $A$ be a clone and let $R$ be a type. Let $R_A$ be the smallest set such that if $r \in R$ is an $n$-ary symbol  and $a_1, ..., a_n \in A$ then $r(a_1, ...., a_n) \in R_A$.

Define $R_A \times A^{\N} \ar R_A$ such that \[(r[a_1, ..., a_n])[b_1, b_2, ... ] = r[a_1[b_1, b_2, ...], ..., a_n[b_1, b_2, ...]].\] Then $R_A$ is a locally finite right algebra over $A$.

The locally finite right algebra $R_A$ has the following universal property.

\lm{Suppose $Q$ is a right algebra over $A$. Suppose $\phi: R \ar Q$ is a function such that for each $n$-ary  $r \in R$, the element $\phi(r)$ has a finite rank $n$. Then $\phi$ extends uniquely to a homomorphism of right algebras over $A$ from $R_A$ to $Q$.}

Let $A$ be a clone and let $T$ be a right algebra over $A$.

Let $\mP(T)$ be the smallest set such that

(i) $T \subset \mP(T)$.

(ii) If $p, q \in \mP(T)$ then $p \wedge q, \neg p, \forall p \in \mP(T)$.

Definite $\mP(T) \times A^{\N} \ar \mP(T)$ inductively on $p \in \mP(T)$ for any $a_1, a_2, ... \in A$:

(i) If $p \in T$ then $p[a_1, a_2, ... ] \in T$ as $T$ is a right algebra over $A$.

(ii) $(\neg p)[a_1, a_2, ...] = \neg (p[a_1, a_2, ...])$.

(iii) $(p \wedge q)[a_1, a_2, ...] = p[a_1, a_2, ...] \wedge q[a_1, a_2, ...]$.

(iv) $(\forall p)[a_1, a_2, ...] = \forall (p[x_1, a_1^+, a_2^+, ...])$.

Then $\mP(T)$ is a predicate algebra over $A$.

\lm{ The predicate algebra $\mP(T)$ over $A$ is locally finite iff the right algebra $T$ is locally finite.}

The predicate algebra $\mP(T)$ has the following universal property:

\lm{Suppose $Q$ is any predicate algebra over $A$. Suppose $\phi: T \ar Q$ is a homomorphism of right algebras over $A$. Then there is a unique homomorphism from $\mP(T)$ to $Q$ extending $\phi$.}

A \la{first-order language} is a pair $\mL = (F, R)$ consisting of a \la{function type}  $F$ and a \la{predicate type $R$}; if an element $e$ of rank $2$ of $R$ is specified then we say $\mL$ is a first-order language with equality. The predicate algebra $P(\mL) = \mP(R_{F(X)})$ is called the \la{first-order algebra for $\mL$}.

\ex{ $\mL_S = (\emptyset, \{e, \in\})$ is the language of set theory. It has a binary predicate symbol $\in$ and a binary equality symbol $e$, with no function symbol.}

\ex{Let $F_a = (\bf 0,$$ \ ', +, \cdot)$ be the arithmetic type with arities $(0, 1,  2, 2)$. Then the first-order language $\mL_a = (F_a, \{e\})$ with equality $e$ is called the \la{language of arithmetic}.
}

Let $\mL = (F, R)$ be a first-order language.

A \la{structure} $D = (D, \{f^D\}, \{r^D\})$ of $\mL$ consists of the following ingredients:

(i) A non-empty set $D$, called the \la{domain of the structure}.

(ii)  For each function symbol $f \in F$ an assignment of an
$n$-ary operation $f^D: D^n \ar D$.

(iii) For each predicate symbol $r \in R$ an assignment
of an $n$-ary relation $r^D: D^n \ar 2$. If $R$ has an equality $e$ we
assume $e[d_1, d_2, ..] =1$ iff $d_1 = d_2$.

Suppose $D$ is a structure of $\mL$.

We first define a function $F(X) \times D^{\N} \ar D$
inductively:

(i) $x_i[d_1, d_2, ...] = d_i$ for any $d_1, d_2,
... \in D^{\N}$.

(ii) $f(t_1, t_2, ..., t_n)[d_1, d_2, ...] =
f^D(t_1[d_1, d_2, ...], t_2[d_1, d_2, ...], ..., t_n[d_1, d_2,
...])$.

Then $D$ is a left $F(X)$-algebra.

We define a function $\alpha: P(\mL) \times
D^{\N} \ar 2$ inductively: for any $r \in R$, $p, q \in P(\mL)$, $t_1, t_2, ...  \in F(X)$ and $d_1, d_2, ... \in D$ let

(i) $r(t_1, t_2, ..., t_n)[d_1, d_2, ...] = r^D[t_1[d_1, d_2, ...], t_2[d_1, d_2, ...], ...]$.

(ii)  $(\neg p)[d_1, d_2, ...] = \neg (p[d_1, d_2, ...]$.

(iii)   $(p \wedge q)[d_1, d_2, ...] = p[d_1, d_2, ...] \wedge q[d_1, d_2, ...]$.

(iv)  $(\forall p)[d_1, d_2, ...] = 1$ if and only if $p[d, d_1,
d_2, ...] = 1$ for any $d \in D$.

\lm{A structure $D$ of $\mL$ determines a classical model  $(D, \alpha)$ for the predicate algebra
$P(\mL)$. Conversely, any classical model for the predicate algebra $P(\mL)$ arises in this way.
}

\section{Fundamental Theorems}
Let $P$ be a predicate algebra over a clone $A$.

A truth valuation $V$ of the proposition algebra $P$ is called a \la{perfect valuation} if the following condition is satisfied:

(C1) If $p \in P$ then $\forall p \in V$ iff $p[a, x_1, x_2, ...] \in V$ for every $a \in A$.

If $V$ is a perfect valuation  we define a multiplication $\alpha: P \times A^{\N} \ar 2$ by \[p[a_1, a_2, ...] = 1 \Leftrightarrow p[a_1, a_2, ...] \in V.\] Then $(A, \alpha)$ is a classical model for $P$ and  $\mV(A) = \{p \in P \ | \ p[x_1, x_2, ...] = p \in V\}$. So $V$ is a local valuation of $P$.

Denote by $P_0$ (resp. $A_0$) the set of closed elements of $P$ (resp. $A$). Then $P_0$ is a Boolean subalgebra of $P$.
A truth valuation $V$ of the proposition algebra $P_0$ is called a \la{closed perfect valuation} if the following condition is satisfied:

(C2) If $p$ is an element of rank $1$ of $P$ then $\forall p \in V$ iff $p[a] \in V$ \ for every closed element $a \in A$.

If $V$ is a closed perfect valuation of $P$  we define a multiplication $\alpha: P \times A_0^{\N} \ar 2$  by \[p[a_1, a_2, ...] = 1 \Leftrightarrow  p[a_1, a_2, ...] \in V \ \text{for any} \ a_1, a_2, ... \in A_0.\] Then $(A_0, \alpha)$ is a classical model for $P$ and  $V$ is the closed valuation determined by $(A_0, \alpha)$.

A maximal local filter $F$ of $P$ is called a \la{perfect filter} if the following condition is satisfied:

(C3) If $p \in P$ and $\neg (\forall p) \in F$ then $(\neg p)[a, x_1, x_2, ...] \in F$ for some $a \in A$.

A maximal proposition filter $F$ of the Boolean algebra $P_0$ is called a \la{closed perfect filter} if the following condition is satisfied:

(C4) If $p \in P$ is an element of rank $1$ and $\neg (\forall p) \in F$ then $(\neg p)[a, x_1, x_2, ...] \in F$ for some $a \in A_0$.

\lm{1. Any perfect valuation is a local valuation.

2, Any closed perfect valuation is a closed valuation.

3. Any perfect filer is a perfect valuation.

4. Any closed perfect filer is a closed perfect valuation. }

\te{\label{te:com} (Completeness Theorem for Predicate Algebras) Suppose $A$ is a locally finite clone. Suppose $P$ is a locally finite predicate algebra over $A$.

1. Any local filter (resp. global filter) $F$ of $P$ is the intersection of all the local valuations (resp. global valuations) of $P$ containing $F$.

2. $T \models_l S$ iff $T \vdash_l S$ for any subsets $T, S$ of $P$.

3. $T \models_g S$ iff $T \vdash_g S$ for any subsets $T, S$ of $P$.
}

\te{(Representation Theorem for Quantifier Algebras) Suppose $A$ is a locally  finite clone and $P$ is a locally finite quantifier algebra over $A$.  Let $A(P)$ be the free left $A$-algebra over the basis $P$. Then

1. $P$ is isomorphic to a subalgebra of a power of $\mP_2(A(P))$.

2. $P$ is simple iff it is isomorphic to a subalgebra of $\mP_2(A(P))$.
}

Let $F_a = (\bf 0,$$ \ ', +, \cdot)$ be the arithmetic type with arities $(0, 1,  2, 2)$. Let $\mL_a = (F_a, \{e\})$ be the language of arithmetic. The first-order algebra $P_a = \mP(\mL_a)$  is called the \la{arithmetic algebra}.

If $P$ is any predicate algebra over $F_a(X)$ with an equality $e$ let  $N_P$ be the subset of $P$ consisting of the following  elements:

(S1) $\neg (e[\bf 0 $, $ x_1'])$.

(S2) $e[x_1', x_2'] \rightarrow e[x_1, x_2]$.

(S3) $ e[x_1 + \bf 0$, $ x_1]$.

(S4) $ e[x_1 + x_2', (x_1 + x_2)']$.

(S5) $e[x_1.\bf 0 $, $ \bf 0]$.

(S6) $e[x_1.(x_2)' , (x_1.x_2) + x_1]$.

(S7) $(p[\bf 0$$] \wedge (\forall(p[x_1] \rightarrow p[x_1']))) \rightarrow \forall p[x_1]$ for any $p \in P$.

A \la{Peano algebra} is a quantifier algebra $P$ with equality $e$ over $F_a(X)$ which is generated by $e$ such that $N_P = \{1\}$.

\te{(Incompleteness Theorem for Peano Algebras). Assume $N_{P_a}$ is consistent. Then
$N_{P_a}$ is not complete (or equivalently, there is a Peano algebra which is not simple).
}

The proofs of these fundamental theorems will be given in subsequent papers.

\end{document}